\documentclass[12pt]{article}
\usepackage{epsfig}
\def\be{\begin{equation}}
\def\ee{\end{equation}}
\def\bea{\begin{eqnarray}}
\def\eea{\end{eqnarray}}
\usepackage{graphicx}

\catcode`\@=11
\def\lsim{\mathrel{\mathpalette\@versim<}}
\def\gsim{\mathrel{\mathpalette\@versim>}}
\def\@versim#1#2{\vcenter{\offinterlineskip
\ialign{$\m@th#1\hfil##\hfil$\crcr#2\crcr\sim\crcr } }}
\catcode`\@=12

\catcode`@=12
\begin{document}
\thispagestyle{empty}
\begin{flushright}
UCRHEP-T550\\
June 2015\
\end{flushright}
\vspace{0.4in}
\begin{center}
{\LARGE \bf Derivation of Dark Matter Parity\\ 
from Lepton Parity\\}
\vspace{0.8in}
{\bf Ernest Ma\\}
\vspace{0.2in}
{\sl Physics \& Astronomy Department and Graduate Division,\\ 
University of California, Riverside, California 92521, USA\\}
\vspace{0.2in}
{\sl HKUST Jockey Club Institute for Advanced Study,\\ 
Hong Kong University of Science and Technology, Hong Kong, China\\}
\end{center}
\vspace{0.8in}

\begin{abstract}\
It is shown that in extensions of the standard model of quarks and leptons 
where additive lepton number $L$ is broken by two units, so that $Z_2$ lepton 
parity, i.e. $(-1)^L$ which is either even or odd, remains exactly conserved, 
there is the possibility of stable dark matter without additional symmetry.  
This applies to many existing simple models of Majorana neutrino mass with 
dark matter, including some radiative models.  Several 
well-known examples are discussed.  This new insight leads to the 
construction of a radiative Type II seesaw model of neutrino mass with 
dark matter where the dominant decay of the doubly charged Higgs boson 
$\xi^{++}$ is into $W^+ W^+$ instead of the expected $l^+_i l^+_j$ lepton 
pairs for the well-known tree-level model.
\end{abstract}

\newpage
\baselineskip 24pt

The origin of neutrino mass has been a fundamental theoretical issue for 
many years.  It is not yet known experimentally whether it is 
Dirac so that an additive lepton number $L$ is conserved, or Majorana 
so that $L$ is broken to $(-1)^L$, i.e. lepton parity which is either 
even or odd, which remains conserved.  Theoretically, it is usually assumed 
to be Majorana, i.e. self-conjugate, and comes from physics at an energy 
scale higher than that of electroweak symmetry breaking of order 100 GeV.  
As such, the starting point of any theoretical discussion of the 
underlying theory of neutrino mass is the effective dimension-five 
operator~\cite{w79,m98}
\begin{equation}
{\cal L}_5 = - {f_{ij} \over 2 \Lambda} (\nu_i \phi^0 - l_i \phi^+) 
(\nu_j \phi^0 - l_j \phi^+) + H.c.,
\end{equation}
where $(\nu_i,l_i), i=1,2,3$ are the three left-handed lepton doublets of 
the standard model (SM) and $(\phi^+,\phi^0)$ is the one Higgs scalar 
doublet.  As $\phi^0$ acquires a nonzero vacuum expectation value 
$\langle \phi^0 \rangle = v$, the neutrino mass matrix is given by
\begin{equation}
{\cal M}^\nu_{ij} = {f_{ij} v^2 \over \Lambda}.
\end{equation}
Note that ${\cal L}_5$ breaks lepton number $L$ by two units.

Consider first the most well-known model where neutrino mass just comes 
from the canonical (Type I) seesaw mechanism with a massive Majorana $\nu_R$. 
The new terms in the Lagrangian are $f \bar{\nu}_R \nu_L \phi^0$ and 
$(M/2) \nu_R \nu_R$.  Hence lepton parity is conserved with $\nu_R$ odd. Now 
consider the simplest possible model of 
dark matter~\cite{sz85} with the addition of just one real singlet scalar 
particle $s$ with odd $Z_2$ dark parity.  How is this linked to lepton 
parity?  The answer is very simple.  Here lepton parity is odd for $\nu$ 
and $l$.  If $s$ is added, then to forbid the $s \nu_R \nu_R$ coupling, 
$s$ must also be odd.  Thus in this simplest model, dark parity is 
identical to lepton parity.  The same holds true if $s$ is replaced by 
a scalar doublet $(\eta^+,\eta^0)$~\cite{dm78}, because its lepton 
parity must also be odd to forbid the $\bar{\nu}_R \nu_L \eta^0$ term.
Suppose a neutral singlet fermion $N_R$ is added to the SM and assumed to be 
dark matter, then to forbid the coupling $\bar{N}_R \nu_L \phi^0$, its 
lepton parity must now be even, but its dark parity should be odd.  In all 
cases, the formula for dark parity is then just $(-1)^{L+2j}$, 
where $j$ is the intrinsic spin of the particle.
At this point, it becomes obvious that this is completely analogous to 
the well-known $R$ parity of supersymmetry, which also stabilizes dark matter.

In recent years, the notion that the underlying physics which generates 
${\cal L}_5$ may be connected to dark matter has motivated a large number 
of studies.  In most cases, an exactly conserved discrete $Z_2$ symmetry 
is imposed for the stability of dark matter, which appears to be unrelated 
to any existing symmetry of the standard model.  As shown above, this is 
actually not the case.  This dark $Z_2$ parity is in fact derivable from 
lepton parity, as discussed in the following three examples.

Consider first the simplest such model of radiative neutrino mass~\cite{m06} 
through dark matter, called ``scotogenic'' from the Greek 'scotos' meaning 
darkness.  The one-loop diagram is shown in Fig.~1.
\begin{figure}[htb]
\vspace*{-3cm}
\hspace*{-3cm}
\includegraphics[scale=1.0]{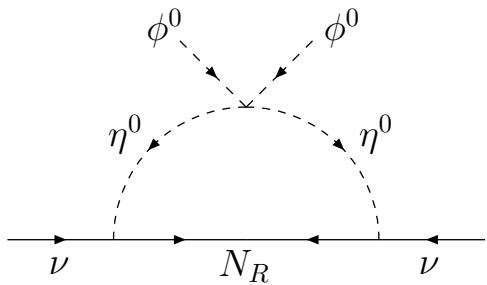}
\vspace*{-21.5cm}
\caption{One-loop $Z_2$ scotogenic neutrino mass.}
\end{figure}
The new particles are a second scalar doublet $(\eta^+,\eta^0)$ and three 
neutral singlet fermions $N_R$.  The imposed $Z_2$ is odd for 
$(\eta^+,\eta^0)$ and $N_R$, whereas all SM particles are even. 
Under lepton parity with $\nu_L$ odd, the same Lagrangian 
is obtained with $\eta$ odd and $N_R$ even.  The imposed dark parity is thus 
again $(-1)^{L+2j}$. 
If the conventional lepton parity assignment is made for $N_R$, i.e. odd,  
then it appears that the model has two $Z_2$ symmetries, but there is 
actually only one as shown above, because dark parity is derivable from 
the new assignment of lepton parity by virtue of the intrinsic spin 
of the new particles.

Consider next the three-loop model~\cite{knt03} with the diagram shown 
in Fig.~2.
\begin{figure}[htb]
\vspace*{-0.5cm}
\hspace*{-0.7cm}
\includegraphics[scale=0.8]{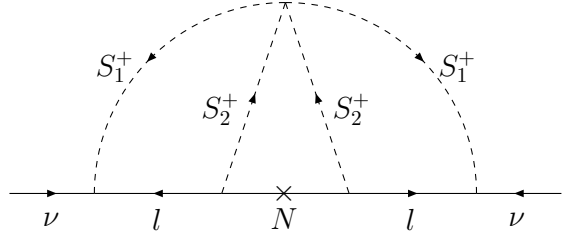}
\vspace*{-18.5cm}
\caption{Three-loop neutrino mass.}
\end{figure}
The new particles are the $S_1^+$, $S_2^+$ scalar singlets, and the $N_R$ 
singlet fermions, where $S_2^+$ and $N_R$ 
are odd.  Using lepton parity with $\nu$, $l$, $S_2^+$ odd and $N_R$, $S_1^+$ 
even, the same Lagrangian is obtained.  Again, dark parity is $(-1)^{L+2j}$.

Another three-loop model~\cite{aks09} has the diagram 
shown in Fig.~3.
\begin{figure}[htb]
\vspace*{-0.5cm}
\hspace*{-1.8cm}
\includegraphics[scale=0.9]{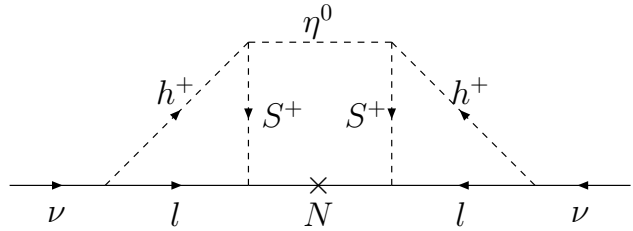}
\vspace*{-21.5cm}
\caption{Another three-loop neutrino mass.}
\end{figure}
The new particles are a second Higgs doublet $(h^+,h^0)$, a neutral scalar 
singlet $\eta^0$, and a charged scalar singlet $S^+$, together with $N_R$, 
where $\eta^0$, $S^+$, and $N_R$ are odd.  Using lepton parity with 
$\nu$, $l$, $S^+$, $\eta^0$ odd, and $h$, $N_R$ even, the same Lagrangian 
is again obtained with dark parity given by $(-1)^{L+2j}$.


There are also models of radiative neutrino mass with larger dark symmetries, 
such as $Z_3$ and $U(1)_D$.  What role does lepton parity play in these 
cases?  The obvious answer is that it cannot generate these symmetries, 
but the question is what happens to the lepton parity itself?  To 
understand this, consider the $Z_3$ dark matter model~\cite{m08} 
of neutrino mass as shown in Fig.~4.
\begin{figure}[htb]
\vspace*{-2cm}
\hspace*{1.2cm}
\includegraphics[scale=0.6]{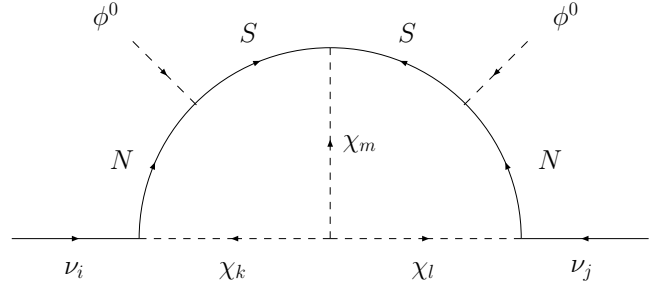}
\vspace*{-10.5cm}
\caption{Two-loop neutrino mass.}
\end{figure}
The new particles are three scalars $\chi$ transforming as $\omega$ under 
$Z_3$, one Dirac fermion doublet $(N,E)$ and one Dirac fermion singlet $S$ 
transforming as $\omega^2$, where $\omega^3 =1$.  To be consistent with 
the Lagrangian of this model, the lepton parity assignment has to be 
odd for $\nu$, $N$ and $S$, which means that the derived dark parity for 
all the new particles are even.  In other words, there is no dark parity 
at all.  The symmetry which stabilizes the new particles is the imposed 
$Z_3$.

Consider now the dark $U(1)_D$ case~\cite{mpr13} as shown in Fig.~5.
\begin{figure}[htb]
\vspace*{-3cm}
\hspace*{-3cm}
\includegraphics[scale=1.0]{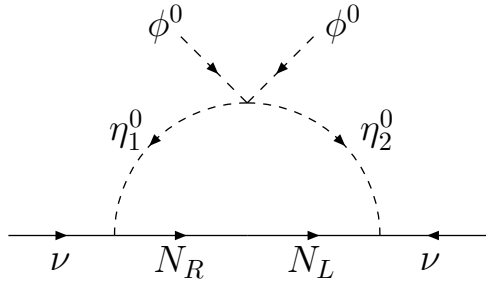}
\vspace*{-21.5cm}
\caption{One-loop $U(1)_D$ scotogenic neutrino mass.}
\end{figure}
The new particles are two scalar doublets $(\eta_1^+,\eta_1^0)$ and 
$(\eta_2^+,\eta_2^0)$ transforming oppositely under $U(1)_D$, and 
three Dirac fermion singlets $N$ transforming as $\eta_1$.  Using 
lepton parity with $\nu$, $\eta_1$, $\eta_2$ odd, a residual $Z_2$ 
symmetry of $U(1)_D$ is obtained.  As expected, the full symmetry 
cannot be reproduced.  For example, the $(\Phi^\dagger \eta_1)^2$ 
term is allowed by lepton parity but not $U(1)_D$.

Since ${\cal L}_5$ is a dimension-five operator, any loop realization is 
guaranteed to be finite.  Suppose a Higgs triplet $(\xi^{++},\xi^+,\xi^0)$ 
is added to the SM, then a dimension-four coupling $\xi^0 \nu_i\nu_j - 
\xi^+ (\nu_i l_j + l_i \nu_j)/\sqrt{2} + \xi^{++} l_i l_j$ is allowed. 
As $\xi^0$ obtains a small vacuum expectation value~\cite{ms98} from its 
interaction with the SM Higgs doublet, neutrinos acquire small Majorana 
masses, i.e. Type II seesaw.   Is there an analogous radiative mechanism 
in this case?  The answer is yes, using the notion of conserved lepton 
number violated only by two units with soft terms.  This new model is 
discussed below.

Lepton number is imposed on all hard (dimension-four) terms of the 
Lagrangian,  with $\xi$ having $L=0$.  Its main purpose is to forbid 
the $\xi^0 \nu \nu$ term.  The scalar trilinear $\bar{\xi}^0 \phi^0 \phi^0$ 
term is allowed and induces a small $\langle \xi^0 
\rangle$, but $\nu$ remains massless.  Note that this assignment is 
opposite to the well-known Gelmini-Roncadelli (GR) model~\cite{gr81} 
where $\xi$ is assigned $L=-2$, so that $\xi^0 \nu \nu$ is allowed but 
$\bar{\xi}^0 \phi^0 \phi^0$ is forbidden.  In the GR case, lepton number is 
spontaneously broken, so the neutrinos acquire mass together with the 
appearance of a massless Goldstone boson (majoron) which has long since 
been ruled out experimentally by $Z$ decay.  In the present model, 
neutrinos will acquire radiative masses with the explicit soft 
breaking of $L$ to $(-1)^L$.  This may be accomplished by adding 
a new Dirac fermion doublet $(N,E)$ with $L=0$, together with 
three complex neutral scalar singlets $s$ with $L=1$.  The resulting 
one-loop diagram is shown in Fig.~6.
\begin{figure}[htb]
\vspace*{-3cm}
\hspace*{-3cm}
\includegraphics[scale=1.0]{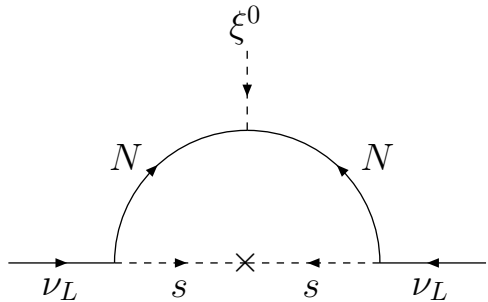}
\vspace*{-21.5cm}
\caption{One-loop neutrino mass from $L=0$ Higgs triplet.}
\end{figure}
Note that the hard terms $\xi^0 N N$ and $s \bar{\nu}_L N_R$ are allowed by 
$L$ conservation, whereas the $s s$ terms break $L$ softly by two units to 
$(-1)^L$.  Note again that the hard term $\xi^0 \nu \nu$ is forbidden, 
or else the usual tree-level Type II seesaw mechanism would have prevailed.
Again, $(N,E)$ and $s$ are odd under $(-1)^{L+2j}$.  The three $s$ scalars 
are the analogs of the three right-handed sneutrinos in supersymmetry, and 
$(N,E)_{L,R}$ are the analogs of the two higgsinos. However, their 
interactions are simpler here and less constrained.  The lighest $s$ is 
a possible dark-matter candidate~\cite{cskw13}, although it is highly 
constrained~\cite{fpu15} from present data if it decouples from all other 
particles except the SM Higgs.  Furthermore, from the allowed 
$(s^*s)(\Phi^\dagger \Phi)$ interactions, electroweak baryogensis~\cite{mrm12} 
may be realized.  Note that whereas one $s$ cannot be both dark matter and 
induce a first-order phase transition in the Higgs potential, as shown 
in Ref.~\cite{ck13}, there are three complex singlets here with mass 
splitting between the real and imaginary parts.  The lightest one is dark 
matter, but the other five may have strong enough couplings to the Higgs boson 
with $CP$ violation to allow successful baryogenesis.  These large loop-induced 
deviations of the Higgs self couplings are presumably observable at a 
future $e^+ e^-$ accelerator for precision Higgs measurements.

The usual understanding of the Type II seesaw mechanism is that the scalar 
trilinear term $\mu \xi^\dagger \Phi \Phi$ breaks lepton number $L$ by two 
units and a small vacuum expectation value $\langle \xi^0 \rangle = u$ may 
be obtained if either $\mu$ is small or $m_\xi$ is large or both.  
More precisely, consider the scalar potential of $\Phi$ and $\xi$.  
\begin{eqnarray}
V &=& m^2 \Phi^\dagger \Phi + M^2 \xi^\dagger \xi + {1 \over 2} \lambda_1 
(\Phi^\dagger \Phi)^2 + {1 \over 2} \lambda_2 (\xi^\dagger \xi)^2 
+ \lambda_3 |2 \xi^{++} \xi^0 - \xi^+ \xi^+|^2 \nonumber \\ 
&+& \lambda_4 (\Phi^\dagger \Phi)(\xi^\dagger \xi) + {1 \over 2} \lambda_5 
[ |\sqrt{2} \xi^{++} \phi^- + \xi^+ \bar{\phi}^0|^2 + 
|\xi^{+} \phi^- + \sqrt{2} \xi^0 \bar{\phi}^0|^2] \nonumber \\ 
&+& \mu (\bar{\xi}^0 \phi^0 \phi^0 + \sqrt{2} \xi^- \phi^0 \phi^+ + 
\xi^{--} \phi^+ \phi^+) + H.c. 
\end{eqnarray}
Let $\langle \phi^0 \rangle = v$, then the conditions for the minimum of $V$ 
are given by~\cite{ms98}
\begin{eqnarray}
m^2 + \lambda_1 v^2 + (\lambda_4 + \lambda_5) u^2 + 2 \mu u &=& 0, \\ 
u[M^2 + \lambda_2 u^2 + (\lambda_4 + \lambda_5) v^2] + 
\mu v^2 &=& 0.
\end{eqnarray}
For $\mu \neq 0$ but small, $u$ is also naturally small because it is 
approximately given by
\begin{equation}
u \simeq {-\mu v^2 \over M^2 + (\lambda_4 + \lambda_5) v^2},
\end{equation}
where $v^2 \simeq -m^2/\lambda_1$.  The physical masses of the $L=0$ Higgs 
triplet are then given by
\begin{eqnarray}
m^2(\xi^0) &\simeq& M^2 + (\lambda_4 + \lambda_5) v^2, \\ 
m^2(\xi^+) &\simeq& M^2 + (\lambda_4 + {1 \over 2}\lambda_5) v^2, \\ 
m^2(\xi^{++}) &\simeq& M^2 + \lambda_4 v^2.
\end{eqnarray}
Since $m_\nu = f_\nu u$, where $f_\nu$ is a Yukawa coupling, there are 
two strategies for making $m_\nu$ small.  (I) One is to keep $f_\nu$ not too 
small, say $f_\nu \sim 0.1$, but make $u \sim 1$ eV.  This implies a very 
small $\mu$ unless $M$ is very large.  For $M$ of order $v$ so that the Higgs 
scalar triplet may be observable, $\mu \sim 1$ eV is required.  To understand 
this small $\mu$ value, one approach is to ascribe it to the breaking of 
lepton number from extra dimensions~\cite{mrs00}.  Another approach is to 
forbid the $\mu$ term at tree level and generate it in one loop~\cite{ks12}.  
(II) The other strategy is to keep $u$ not too small, say 0.1 GeV, but make 
$f_\nu$ very small.  This is what happens here because the $L=0$ assignment 
for $\xi$ means $f_\nu = 0$ at tree level.   It is then generated in one 
loop as shown in Fig.~6.  Let the relevant Yukawa interactions be given by
\begin{equation}
{\cal L}_Y = f_s s \bar{\nu}_L N_R + {1 \over 2} f_R \xi^0 N_R N_R + 
{1 \over 2} f_L \xi^0 N_L N_L + H.c.,
\end{equation}
together with the allowed mass terms $m_E(\bar{N} N + \bar{E} E)$, $m_s^2 
s^*s$, and the $L$ breaking soft term $(1/2)(\Delta m_s^2)s^2 + H.c.$, then 
\begin{equation}
m_\nu = {f_s^2 u r x \over 16 \pi^2} [f_R F_R(x) + f_L F_L(x)],
\end{equation}
where $r = \Delta m_s^2/m_s^2$ and $x=m_s^2/m_E^2$, with
\begin{eqnarray}
F_R(x) &=& {1+x \over (1-x)^2} + {2 x \ln x \over (1-x)^3}, \\ 
F_L(x) &=& {2 \over (1-x)^2} + {(1+x) \ln x \over (1-x)^3}. 
\end{eqnarray}
Since $m_\nu$ is now suppressed relative to $u$, the latter value may be as 
large as 0.1 GeV, using for example $x \sim f_R \sim f_L \sim 0.1$, $r \sim 
f_s \sim 0.01$.  This implies that $\xi$ may be as light as a few 
hundred GeV and be observable, with $\mu \sim 1$ GeV.  Note that  $u \sim 0.1$ 
GeV has negligible contribution (of order $10^{-6}$) to the precisely 
measured $\rho$ parameter $\rho_0 = 1.00040 \pm 0.00024$~\cite{pdg14}. 
For $f_s \sim 0.01$ and $m_E$ a few hundred GeV, the new contributions 
to the anomalous muon magnetic moment and $\mu \to e \gamma$ are also 
negligible in this model.

As for the decay of $\xi$, its effective couplings to leptons are now 
very small, unlike the tree-level Type II seesaw model, where the decay 
of $\xi^{++}$ to same-sign dileptons is expected to be dominant.  Current 
experimental 
limits~\cite{atlas14} on the mass of $\xi^{++}$ into $e\mu$, $\mu \mu$, 
and $ee$ final states are about 490 to 550 GeV, assuming for each a 
100\% branching fraction.  These limits are not valid in the present model. 
Instead, $\xi^{++} \to W^+ W^+$ should be considerd~\cite{kkyy14}, for which 
the present limit on $m(\xi^{++})$ is only about 84 GeV~\cite{kkyy15}.  
A dedicated search of the $W^+ W^+$ mode in the future is clearly called for.

If $m(\xi^{++}) > 2 m_E$, then the decay channel $\xi^{++} \to E^+ E^+$ 
opens up and will dominate.  In that case, the subsequent decay 
$E^+ \to l^+ s$, i.e. charged lepton plus missing energy, will be the 
signature.  The present experimental limit~\cite{cms14} on $m_E$, 
assuming electroweak pair production, is about 260 GeV if $m_s < 100$ GeV 
for a 100\% branching fraction to $e$ or $\mu$, and no limit if $m_s > 100$ 
GeV.  There is also a lower threshold for $\xi^{++}$ decay, i.e. $m(\xi^{++})$ 
sufficiently greater than $2 m_s$, for which $\xi^{++}$ decays through a 
virtual $E^+ E^+$ pair to $s s l^+ l^+$, resulting in same-sign dileptons plus 
missing energy.

The lepton number symmetry $L$ may be promoted to the well-known $B-L$ 
gauge symmetry, but then three neutral singlet fermions $\nu_R$ 
transforming as $-1$ under $U(1)_{B-L}$ are usually added to satisfy 
the anomaly-free conditions.  This means that neutrinos obtain tree-level 
Dirac masses from the allowed term $\bar{\nu}_L \nu_R \bar{\phi}^0$, and 
Type II seesaw would not be necessary.  However, another possibility 
for anomaly cancellation is to have the three $\nu_R$'s tranform as 
$(-4,-4,5)$~\cite{mp09,ms15}.  In that case, $\bar{\nu}_L \nu_R \bar{\phi}^0$ 
is forbidden, and the spontaneous breaking of $U(1)_{B-L}$ leads to 
a radiative Type II seesaw model as described.
 
Finally, suppose the generation of neutrino mass is extended to (some) 
quarks and charged leptons through dark matter~\cite{m14}, then lepton 
parity may be promoted to matter parity, i.e. $(-1)^{3B+L}$ and dark 
parity becomes exactly $R$ parity, i.e. $(-1)^{3B+L+2j}$, in complete 
analogy with what happens in the minimal supersymmetric standard model 
(MSSM).  Of course, here the usual tree-level Yukawa couplings of the 
Higgs doublet to quarks and charged leptons must be forbidden by an 
imposed flavor symmetry~\cite{m14}.

In conclusion, it has been pointed out in this paper the very simple idea 
that for many well-studied models of neutrino mass with dark matter, lepton 
parity, if appropriately defined, leads automatically to dark parity. 
Several specific examples of existing models were discussed, together with 
a new radiative Type II seesaw model of neutrino mass with dark matter, 
which predicts a doubly charged Higgs boson with the interesting 
dominant decay mode of $\xi^{++} \to W^+ W^+$~\cite{kkyy14,kkyy15}, 
below the threshold of the production of dark matter.

\medskip
This work is supported in part 
by the U.~S.~Department of Energy under Grant No.~DE-SC0008541.

\bibliographystyle{unsrt}

\end{document}